\documentclass [12pt]{article}
\usepackage{latexsym}
\usepackage{amsfonts}
\usepackage{amsmath}
\usepackage{acronym}
\newcommand{\be}{\begin{equation}}
\newcommand{\ee}{\end{equation}}
\newcommand{\bea}{\begin{eqnarray}}
\newcommand{\eea}{\end{eqnarray}}
\makeatletter
\@addtoreset{equation}{section}

 \makeatother
\begin{document}
\normalsize
\title{Interpreting Dirac variables in terms of the Hilbert space of gauge-invariant and Poincare-covariant states.}
\author
{{\bf L.~D.~Lantsman}\\  
Tel.  049-0381-799-07-24,\\
llantsman@freenet.de}
\maketitle
\begin {abstract}
The goal of this note is to give a description of  Dirac variables in Abelian as well as non-Abelian gauge models in terms of gauge-invariant and Poincare-covariant states sweeping a Hilbert space ${\cal H}_{\rm vac}$ 

The next our conjecture concerns the spontaneous breakdown of the Abelian $U(1)$ symmetry in the `discrete` $U(1)\to {\bf Z}$ wise. We suppose that  gauge charges are preserved in this case. 
\end{abstract}
\newpage 
\tableofcontents
\newpage
\section{Introduction.}
The topic 'Dirac variables' becomes important  when  the so-called {\it fundamental}	quantization of gauge theories \cite{Dir} is  in the question.
Briefly speaking, this fundamental	quantization method comes to the reduction of the  appropriate Hamiltonians in terms of {\it physical}, i.e.  always transverse and gauge invariant, variables.

 Dirac variables are just  such variables. There are physical fields which are solutions to the {\it Gauss law constraint} \footnote{ Generally speaking, a constraint equation in the Hamiltonian formalism  can be defined as following \cite{LP1}. Constraint equations relate initial data for
spatial components of the fields involved in a (gauge) model to initial data of their temporal components.

The Gauss law constraint has an additional specific that it is simultaneously the one of equations of motion (to solve these, it is necessary a measurement of initial
data \cite{LP1}). This is correctly for QED as well for non-Abelian theories (in particular, for QCD), i.e. for the so-called {\it particular} theories involving \cite{Gitman} the singular  Hessian matrix
$$ M_{ab}= \frac {\partial ^2 L}{\partial \dot q ^a\partial\dot q ^b} $$
(with $L$ being the Lagrangian of the studied theory,   $q ^i$ being the appropriate degrees of
freedom and $\dot q^i$ being their time derivatives).

The Hessian matrix $M$ becomes singular (i.e. ${\rm det} M=0$), sinse the identity 
$$ \partial  L/ \partial  \dot A_0 \equiv 0$$
which 'temporal' components $A_0$ of gauge fields always satisfy in particular theories (in other words, particular theories involve zero canonical momenta $\partial  L/ \partial \dot A ^0$ for  temporal components $A_0$ of gauge fields).

Thus temporal components $A ^0$ of gauge fields are, indeed, non-dynamical degrees of freedom
in particular theories, the quantization of which contradicts the Heisenberg uncertainty
principle.
}. 

Dirac \cite{Dir} and, after him, other authors of the first classical studies in quantization
of gauge fields, for instance \cite{Heisenberg,Fermi}, eliminated temporal components of gauge 
fields by gauge transformations. The typical look of such gauge transformations is \cite{David3}
\be \label{udalenie} v^T({\bf x},t)(A_0+\partial_0)(v^{T})^{-1}({\bf x},t)=0. \ee 
This equation may be treated as that specifying the gauge matrices $ v^T({\bf x},t)$. This, in
turn, allows to write down the gauge transformations for spatial components of gauge fields \cite{LP1}  (say, in a non-Abelian gauge theory)
\be \label{per.Diraca} {\hat A}^D_i({\bf x},t):=v^T({\bf x},t)({\hat A}_i+\partial_i)(v^{T})^{-1}({\bf x},t); \quad {\hat A}_i= g \frac {\tau ^a}{2i}A_{ai}.
\ee
It is easy to check that the functionals ${\hat A}^D_i({\bf x},t)$ specified in such a way are gauge invariant and transverse fields: 
\be \label{transvers} \partial_0  \partial_i {\hat A}^D_i({\bf x},t) =0; \quad  u ({\bf x},t)  {\hat A}^D_i({\bf x},t)   u ({\bf x},t)^{-1}= {\hat A}^D_i ({\bf x},t)  \ee
for gauge matrices $ u ({\bf x},t)$. \par 
Following Dirac \cite{Dir}, we shall refer to the functionals ${\hat A}^D_i({\bf x},t)$ as to the {\it Dirac variables}.  
The Dirac variables ${\hat A}^D_i$ may be derived by resolving the Gauss law constraint
\be \label {Gau} \partial W/\partial A_0=0\ee
(with $W$ being the action functional of the considered gauge theory). \par 
Solving Eq. (\ref{Gau}), one expresses  temporal components $A_0$ of gauge
fields $A$ through their  spatial components; by  that the nondynamical components $A_0$ are indeed ruled out from the appropriate Hamiltonians.  
Thus the reduction of particular gauge theories occurs over the surfaces of the appropriate Gauss law constraints. 
Only upon expressing temporal components $A_0$ of gauge
fields $A$ through their  spatial components one can perform gauge transformations (\ref{per.Diraca}) in order to  turn spatial components $\hat A_i$ of gauge
fields into gauge-invariant and transverse Dirac variables ${\hat A}^D_i$. Thus, formally, temporal components $A_0$ of these fields become zero.  
By that the Gauss law constraint (\ref{Gau}) acquires the form  \cite{Pervush2}
\be \label{perpen} \partial_0 \left(\partial_i \hat A^D_i({\bf x},t) \right) \equiv 0.\ee
For  further detailed study of the "technology" getting Dirac variables, in particular gauge theories, we recommend  the works \cite{Pervush2, Nguen2, Azimov} (four-dimensional constraint-shell QED involving electronic currents) and \cite{LP1,LP2} (the Minkowskian non-Abelian Gauss law
constraint-shell model involving vacuum BPS monopole  solutions).\par
\medskip 
Dirac variables prove to be manifestly relativistically covariant.  
  Relativistic properties of Dirac variables in gauge theories were investigated in the papers \cite{Heisenberg} (with the reference
to the unpublished note by von Neumann), and then this job was continued by I. V.
Polubarinov in his review \cite{Polubarinov}. \par
These investigations displayed that there exist such relativistic transformations of Dirac variables that maintain transverse gauges of  fields. 
More precisely,  Dirac variables ${\hat A}^{(0)D}$ observed in a rest  reference frame $\eta_\mu^0=(1,0,0,0)$ in the Minkowski space-time  (thus $\partial_i{\hat A}^{(0)D}_i=0$), in a moving reference frame 
$$ \eta ^{\prime}  = {\eta^0 }+\delta^0_L{\eta_0} $$
are also transverse, but now regarding the new reference frame $\eta ^{\prime}$ \cite{David3,Pervush2},
 \be \label{trp} \partial_\mu{\hat A}^{D\prime}_\mu=0.
\ee
In particular, $A_0(\eta^0) = A_0(\eta^{0\prime}) =0$, i.e. the Dirac removal (\ref{udalenie}) \cite{Dir, David3} of temporal components of gauge fields is transferred from the rest to the moving reference frame. 
In this consideration \cite{Heisenberg,Pervush2,Polubarinov}, $\delta^0_L $  are ordinary total Lorentz transformations of coordinates, involving  appropriate transformations of fields (bosonic and fermionic).  
When one transforms fields entering the  gauge theory into Dirac variables in a rest reference frame $\eta_0$ and then goes over to a moving reference frame $\eta^\prime$, Dirac variables $\hat A^D$, $\psi^D$, $\phi^D$   are suffered relativistic transformations consisting of two therms.

The first item is the response of Dirac variables onto ordinary total Lorentz transformations
of coordinates (Lorentz busts)
$$x'_k=  x_k+ \epsilon_ k t, \quad t'= t+ \epsilon_ k x_k, \quad \vert \epsilon_ k \vert \ll 1. $$
The second therm corresponds to "gauge" Lorentz transformations $\Lambda(x)$ of Dirac variables  $\hat A^D$, $\psi^D$, $\phi^D$   \cite{Pervush2, Nguen2}  (latter two ones are fermionic and scalar fields, respectively) \footnote{It may be demonstrated \cite{Pervush2, Nguen2,Azimov,Arsen} that the  transformations (\ref{per.Diraca}), turning gauge fields $A$ into Dirac variables $\hat A^D$, imply the $\psi^D= v^T({\bf x},t)\psi$ transformations for fermionic fields $\psi$ and $\phi^D= v^T({\bf x},t)\phi$ transformations for spin 0 fields.}:
$$  \Lambda(x)\sim \epsilon_ k \dot A^D_k (x)\Delta^{-1},    $$ 
with 
$$\frac{1}{\Delta}f(x)=-\frac{1}{4\pi}\int d^3y\frac{f(y)}{|\bf{x}-{\bf y}|}
$$
for any continuous function $ f(x)$.    
Thus any relativistic transformation for  Dirac variables may be represented as the sum of two enumerated therms. 
For instance \cite{Pervush2},
\begin{eqnarray}
\label{ltf} 
 A_{k}^{D} [ A_{i} &+& \delta_{L}^{0} A ] - A_{k}^{D} [ A ]  =  \delta_{L}^{0} A_{k}^{D} + \partial_{k} \Lambda,  \end{eqnarray} 
 \begin{eqnarray}
\label{ltf1}
 \psi^{D} [ A &+& \delta_{L}^{0} A , \psi + \delta_{L}^{0} \psi ] -  \psi^{D} [ A, \psi ] = \delta_{L}^{0} \psi^{D} + i e \Lambda  (x^{\prime}) \psi^{D}.  \end{eqnarray}  
 
 \bigskip The aim of the present study is to state the said above about the Dirac variables in the formalized language \cite{Logunov} of the Hilbert space ${\cal H}_{\rm vac}$ of physical states  gauge invariant with respect to the 'large' group $\cal G$ and Poincare covariant simultaniously. Such a construction is suitable to the description of the Dirac variables referred to the vacuum of the quested gauge model quantized by Dirac \cite{Dir}.
 
 But the above construction permits its generalization for nonvacuum states. Examples of such nonvacuum states are so-called {\it multipoles}: perturbation excitations  over the (non-Abelian) vacuum represented by 'monopolelike' solutions. The former possess the same topological numbers that 
the appropriate  monopoles.

 The way to such generalization is \cite{Logunov} to extract the 'small' subgroup ${\cal G}_0$ of Poincare invariant gauge transformations in the 'large' group $\cal G$.  In the QED case such 'small' subgroup ${\cal G}_0$ can be set by the condition 
 \be \label{small1}
 \delta_{L}^{0} A_{k}^{D}=-\partial_{k} \Lambda,
 \ee
 following directly from Eq. (\ref{ltf}) for the appropriate Dirac variables $A^{D}$.
 
 For the  non-Abelian Yang-Mills (further YM) theory quantized by Dirac \cite{Dir} and involving BPS monopole vacuum \footnote{Besides the papers \cite{LP1,David3,Pervush2,LP2}, in which    the above theory was outlined, we recomend also our readers the recent survey \cite{fund}, summarized in an  enough compact shape the maid in this direction in the last ten years. }, it will be argued in the next section that the 'small' subgroup ${\cal G}_0$ coincides with the group of topologically trivial ($n=0$) gauge transformations.
 
 As a consequence  of extracting the 'small' subgroup ${\cal G}_0\in {\cal G}$, the physical Hilbert space $\cal H$ can be decomposed in the unique way into the direct sum of $\bf \Gamma={\cal G}/{\cal G}_0$-orthogonal subspaces ${\cal H}^{(\tau)}\neq 0$:
 \be \label{direct}
 {\cal H}=\oplus _\tau {\cal H}^{(\tau)}.
 \ee
 In particular, we shall show that in the case of the YM model with vacuum BPS monopole solutions quantized by Dirac, $\tau$ runs over the set of integers, the set of topological numbers. 
 
 \medskip In Section 3 of the present study we shall discuss the one interesting result associated with the spontaneous breakdown of the Abbelian $U(1)$ gauge symmetry in the 'discrete'
 \be \label{discret} U(1)\to {\bf Z} \ee
 wise.  
 
 Geometrically, this means that the circle $U(1)\simeq S^1$ disintegrates to its topological sectors; just the isomorphism between these topological sectors and the set $\bf Z$ of integers is expressed in Eq. (\ref{discret}).   
 
 This 'nonordinary' way (\ref{discret}) to break down the (initial) $U(1)$ gauge symmetry was discussed recently in the paper \cite{disc} with the example of  ${\rm He}^4$. In  that case, as it was argued in \cite{disc},  the way (\ref{discret}) to gauge $U(1)$ spontaneous breakdown provides the existence of rectilinear vortices \cite{Halatnikov} in a liquid ${\rm He}^4$  specimen rested in a (narrow) cylindrical vessel and simultaneously superfluid potential motions therein \cite{Kapitza,Landau}. It is easy to show herewith that rectilinear vortices \cite{Halatnikov} inside a  liquid ${\rm He}^4$ specimen contained in a narrow cylindrical vessel possess nonzero  topological numbers $n\neq 0$, while superfluid potential motions in a ${\rm He}^4$ possess namely  the zero  topological number. 
 
 \medskip Our principal result in Section 3 is following. We conjecture that the above isomorphism (\ref{discret}) has as its consequence the result rather opposite to that  stated in \cite{Logunov} (in \S 10.3.B): the authomorphisms group $\gamma_h$ ($\gamma_h\in U(1)$) acted in the field operators algebra $\cal F$ now can be realized by the  group of quasi-unitary operators in the quasi-Hilbert space $\Huge G$ which leave invariant the appropriate vacuum vector $\vert 0>$.

 \section{The vacuum sector of a gauge model permits its description in terms of Dirac variables. }
 Let us now suppose \cite{Logunov} that one has constructed some $*$-algebra (algebra with involution) $\cal U$, which we shall call {\it the observers algebra } henceforth, is built from fundamental (quantum) fields in the quasi-Hilbert space $\Huge G$ \footnote{In difinition \cite{Logunov}, a quasi-Hilbert space $\Huge G$ is a  Frechet space (F-space) possessing a Hilbert topology with an indefinite scalar product $(\Phi,\Psi)$; herewith the indefinite scalar product and the Hilbert topology are coordinated in the following way. Each linear continuous functional $F$ on $\Huge G$ can be represented, in the unique way, as
 $$ F(\Psi)= (\Phi,\Psi), \quad \Psi \in \Huge G,$$
 where $\Phi$ is a vector from $\Huge G$. Vice verse, for each  $\Phi \in \Huge G$,  above Eq. specifies a linear continuous functional $F(\Psi)$ on $\Huge G$.  
  }. The fundamental principle according which the algebra $\cal U$ can be chosen is the gauge invariance. This means that some group $\cal G$ acts with $*$-isomorphisms $A\to\gamma_g(A)$ on the field algebra $\cal F$. In this context, we shal refer to the group $\cal G$ as to the {\it large gauge group}. It is quite naturally herewith to impose the following condition of the ({\it sequentional}) continuity: if $A_n\to A$ on $\cal F$, then $\gamma_g(A_n) \to\gamma_g(A)$ at each $g\in\cal G$. 
  
 \medskip The observables algebra $\cal U$ is specified as a subalgebra of all the $\cal G$-invariant elements in $\cal F$. In order that this algebra $\cal U$ maps in itself under the authomorphisms $\alpha_{(\alpha,\underline \Lambda)}$ of the connected and one-connected Poincare group ${\cal B}_0$ (in this context, $\alpha$ in the brackets stands for the Poincare translations, while ${\underline \Lambda}\in {\rm SL(2,C)}$), we shall suppose that ${\cal B}_0$ acts with authomorphisms on $\cal G$ ($g\to g ^\prime\equiv g^{(\alpha,\underline \Lambda)}$); thus the ${\cal B}_0$-covariance condition
 \be\label{covr}
 \alpha_{(\alpha,\underline \Lambda)}\gamma_g\alpha^{-1}_{(\alpha,\underline \Lambda)}=\gamma_g^\prime
 \ee
 is satisfied.
 
 Besides that, according to the physical sense of observable values, those would be transformed by the one-valued representation of the Lorentz group, therefore we claim that among the authomorphisms $\gamma_g$ ($g\in \cal G$) there exists the transformation $\alpha_{(0,-1)}$ that
 \be\label{one}
 \alpha_{(0,-1)}(A)=A.
 \ee 
  It is easy to see now that Dirac variables (\ref{per.Diraca}) \cite{Pervush2} form such an observables algebra $\cal U$. The additional claim (\ref{trp}) \cite{Pervush2} for those Dirac variables to remain transverse  at Poincare transformations (\ref{ltf}) does not change the picture.
  
 \bigskip The following axiom \cite{Logunov} makes possible the physical interpretation of the stated formalism.
 
 \bigskip {\bf  Axiom 1.} {\it The large gauge group $\cal G$ is specified acting by $*$-authomorphisms authomorphisms on the field algebra $\cal F$, herewith} 
 
 (a) {\it the Poincare covariance condition} (\ref{covr}) {\it is satisfied};
 
 (b) {\it the vacuum functional $<0\vert A\vert 0>$ over the observables algebra $\cal U$} ({\it i.e. over the subalgebra of all the $\cal G$ invariant elements from  $\cal F$ is a lineal positive functional}:
 \be\label{posit}
 <0\vert AA^*\vert 0> \geq 0 \quad {\rm at}\quad A\in {\cal U}.
 \ee
 With the aid of the Gelfand-Naimark-Segal (GNS)  construction \footnote{To set the GNS construction means the following \cite{Logunov}. For a set positive functional $F$ on a $*$-algebra $\cal U$ a (cyclic) representation $\pi_F$ of the algebra $\cal U$ in the given Hilbert space with the cyclic vector $\Phi$ can be specified in such a way that
$$ F(A)=<\Phi_{F},\pi_{F}(A)\Phi_{F}> \quad {\rm at}\quad {\rm all}\quad A\in {\cal U}.$$
The representation $\pi_F$ is specified in this way uniquely to within the unitary equivalence (correlating cyclic vectors of different representations).
},  the vacuum expectation value (as a  positive functional) specifies the cyclic representation $\pi_{\rm vac}$ of the observables algebra $\cal U$ in a Hilbert space $\cal H_{\rm vac}$ with the cyclic (vacuum) vector $\Psi_{\rm vac}$; this representation  is called {\it the vacuum one}.

\bigskip Generalizing the above construction, one can get the non-vacuum states. Let a normal subgroup ${\cal G}_0$, maping in itself under the Poincare ${\cal B}_0$ transformations, be pick out in the  group ${\cal G}$. We shall refer  to this group ${\cal G}_0$ as to the {\it small gauge group}.




\bigskip In this study we shall restricted by the case when the quotient group ${\bf \Gamma}={\cal G}/{\cal G}_0$ is a compact Lee group; let us call it {\it the effective gauge group}.

 By analogy with the definition of the observables algebra, we, following the authors \cite{Logunov}, shall call the set $\cal B$ of all  the ${\cal G}_0$-invariant elements of $\cal F$ {\it the algebra of physical values}. Since $\gamma_g(A)$ (at $g\in {\cal G}$, $A\in \cal B$) depends only on the coset class $h\in \bf \Gamma$ of the element $g$, then, taking liberties, one can write down $\gamma_g(A)=\gamma_h(A)$; by this the group $\bf \Gamma$ acts with $*$-authomorphisms of the algebra $\cal B$.
 
 The next natural proposition consists in the fact that at each $A\in \cal B$ and $\Phi$, $\Psi$ are Schwartz distributions (with compact supports), the functional $<\Phi\vert \gamma_h(A)\vert \Psi>$ is continuous by $h\in \bf \Gamma$ and that one can always write down the average of an element $A\in \cal B$ by the group ${\bf \Gamma}$, $\int \gamma_h(A) dh$, which is simultaneously the element from the observables algebra $\cal U$. Herewith
 \be \label{corel}
 <\Phi\vert \int_{\bf \Gamma}\gamma_h(A) dh\vert \Psi>=\int_{\bf \Gamma}<\Phi\vert \gamma_h(A)\vert \Psi>
 \ee 
 (here $dh$ is the invariant measure on $\bf \Gamma$, and such that $\int_{\bf \Gamma} h=1$).
 
\medskip Thus in order to have the possibility to construct {\it charge sectors} of the  observables algebra $\cal U$, let us assume the following axiom.

{\bf Axiom 2.} (the properties of the physical values algebra). {\it Let the small gauge group ${\cal G}_0$ be given, which is the Poincare invariant normal subgroup in ${\cal G}$ implicating the compact quotient group ${\bf \Gamma}={\cal G}/{\cal G}_0$; herewith}

(a) {\it the authomorphisms $\gamma_h $} ($h\in \bf \Gamma$) {\it of the} $*$-{\it algebra $\cal B$ of physical values} ({\it the latter one is determined as the subalgebra of all the ${\cal G}_0$-invariant elements of ${\cal F}$}) {\it commute with the authomorphisms $\alpha_{a,1}$ of the Poincare group ${\cal B}_0$};

(b) {\it Eq.} (\ref{posit}) {\it determines the averaging operator } ({\it over the group ${\bf \Gamma}$}) {\it from the algebra ${\cal B}$ in ${\cal U}$};

(c) {\it the expression}
\be \label{sa}
s(A)=\int_{\bf \Gamma} <0\vert \gamma_h(A)\vert 0>dh, \quad A\in {\cal B},
\ee
{\it is a lineal positive functional on $\cal B$, i.e.}
\be \label{sa1}
s(AA^*)\geq 0.
\ee

\medskip Again the GNS construction \cite{Logunov} over the lineal positive functional $s(A)$ of  the algebra $\cal B$ allows to construct the cyclic representation $\pi$ of  this algebra $\cal B$ with the cyclic vacuum vector $\Psi_0$ in some (physical) Hilbert space $\cal H$  in such a way that
\be \label{scycl}
s(A)\equiv <\Psi_0,\pi(A)\Psi_0>,\quad A\in {\cal B}.
\ee
It is obvious  that the functional $s(A)$ is ${\cal B}_0$ and ${\bf \Gamma}$-invariant, therefore  there exist unitary  representations $U(a,{\underline\Lambda})$ of  the Poincare group ${\cal B}_0$ and $V(h)$ of  the  group ${\bf \Gamma}$ \footnote{The existence of such unitary  representations is the essence of  the following statement \cite{Logunov}.

 Let $F$ be a positive functional on a $*$-algebra $\cal U$ invariant with respect to an authomorphism $\gamma$ of this algebra $\cal U$, and let $\pi_F$, ${\cal H}_F$, $\Phi_F$ are the components of the GNS construction. Then  there exists the unitary operator $U_F$ in ${\cal H}_F$ that
$$  \pi_F(\gamma(A)) =U_F\pi_F(A)U_F^{-1} \quad {\rm at~~ all~~the }\quad A\in {\cal U}.$$ 
And moreover, if one claim that 
$$ U_F\Phi_F=\Phi_F,$$
then  above   two conditions specify $U_F$ in a unique way.
}; herewith the operators $U(a,{\underline\Lambda})$ commute with the operators $V(h)$, they leave invariant the  vacuum vector $\Psi_0$. Simultaneously,  the subspace of all the translation-invariant vectors in $\cal H$, generally speaking, can have the dimension greater than one (then one speaks about the degeneration  of  the vacuum), while the representation $\pi$ can be reducible. 

\medskip According to the theory of the unitary  representations (see, for instance, \S 1.4.2 in \cite{Vilenkin}), the physical Hilbert space $\cal H$ can be decomposed in a unique way into the direct sum of orthogonal $\bf {\Gamma}$-invariant subspaces ${\cal H}^{(\tau)}\neq \{0\}$. Thus Eq. (\ref{direct}) is satisfied. 

In Eq. (\ref{direct}) the index $\tau$ runs over the some family ${\cal F}^{\prime}$ of mutually nonequivalent (finite-dimensional) unitary irreducible representations of the group $\bf {\Gamma}$, while the subrepresentation of the group $\bf {\Gamma}$ in ${\cal H}$ is unitary equivalent to the representation of the shape $\tau(h)\otimes {\bf 1}$ (herewith ${\cal H}^{(\tau)}$ is isomorphic to the tensor product of the representation space for $\tau$ with some Hilbert space).

\bigskip In the  YM model with vacuum BPS monopole solutions quantized by Dirac \cite{Dir} (the important for us peculiarity of this  model is the presence therein vacuum Higgs modes in the shape of BPS monopoles \cite{LP1,David3,Pervush2,LP2,fund}), above Eq. (\ref{direct}) permits an interesting physical interpretation.

 The cornestone of this   model are so-called {\it topological Dirac variables} got, according to the general 'Dirac scheme' \cite{Dir}, at resolving the appropriate Gauss law constraint (\ref{Gau}), now in terms of YM fields $\hat A$. 
 
 As it was argued in Ref. \cite{David3},  topological Dirac variables in  the  YM theory with vacuum BPS monopole solutions take the shape
 \bea
\label{degeneration}
\hat A_k^D = v^{(n)}({\bf x})T \exp \left\{\int  \limits_{t_0}^t d {\bar t}\hat A _0(\bar t,  {\bf x})\right\}\left({\hat A}_k^{(0)}+\partial_k\right ) \left[v^{(n)}({\bf x}) T \exp \left\{\int  \limits_{t_0}^t d {\bar t} \hat A _0(\bar t,{\bf x})\right\}\right]^{-1}, \eea 
with  the symbol $T$  standing for   time ordering  the matrices under the exponent sign and involving topological numbers $n\in \bf Z$. \par
Topological Dirac variables (\ref{degeneration}) satisfy the transverse gauge (\ref{transvers}).

In the  YM model with vacuum BPS monopole solutions quantized by Dirac \cite{LP1,David3,Pervush2,LP2,fund} this transverse gauge (\ref{transvers}) takes the account of such  vacuum BPS monopoles and (non)trivial topologies:
\be
\label{transv}
D_i^{ab} (\Phi _k^{(n)})\hat{A}^{i(n)}_b =0. 
\ee
In this context $\Phi _k^{(n)}$ are just the {\it Gribov topological copies} of the topologically trivial YM BPS monopole solutions $\Phi _k^{(0)}$ \cite{LP1,LP2,BPS,Gold,Al.S.}.

It can be argued (see, for instance, \cite{rem2}) that in the time instant $t=t_0$ the topological Dirac variables (\ref{degeneration}) (satisfying the transverse gauge (\ref{transv})) acquire the look
\be
\label{degeneration1}
{\hat A}^{(n)}_k= v^{(n)}({\bf x}) ({\hat A}_k^{ (0)}+
\partial _k)v^{(n)}({\bf x})^{-1},\quad v^{(n)}({\bf x})=
\exp [n\Phi _0({\bf x})].
\ee 
Here $\Phi _0({\bf x})$ is the so-called {\it Gribov phase} \cite{LP1,rem2}, that is
\be
 \label{phasis}
{\hat \Phi}_0(r)= -i\pi \frac {\tau ^a x_a}{r}f_{01}^{BPS}(r), \quad 
f_{01}^{BPS}(r)=[\frac{1}{\tanh (r/\epsilon)}-\frac{\epsilon}{r} ].
 \ee
 We see that the Gribov phase $\Phi _0({\bf x})$ is directly proportional to the BPS ansatz $f_{01}^{BPS}(r)$.
 
\medskip Now, with the aid of simple arguments, we shall attempt to show that the trivial topology $n=0$ in the YM model with vacuum BPS monopoles  quantized by Dirac \cite{LP1,David3,Pervush2,LP2,fund} corresponds namelly to Poincare (${\cal B}_0$) invariant YM fields forming the above group ${\cal G}_0$ of physical values. 

This becomes actually obvious since  topologically trivial YM fields can be transformed, by continuous defformations, to their zero values. Latter are manifestly Poincare  invariant.

More strict  arguments are purely mathematical. Such manipulations with topologically trivial Gribov multipliers 
$$v^{(0)}({\bf x})=
\exp [0\cdot\Phi _0({\bf x})] $$
in (\ref{degeneration1}) as taking their derivatives $\partial_k$ or the gauge transformations
$$ {\hat A}^{(0)}_k\to v^{(0)}({\bf x}) {\hat A}_k^{ (0)} v^{(0)}({\bf x})^{-1}$$
leave topologically trivial YM fields ${\hat A}^{(0)}_k$ {\it again topologically trivial}. Really, in the first case any derivative of the BPS ansatz $f_{01}^{BPS}(r)$ should be multiplied by the zero topological charge $n=0$ and this product is zero.  In the second case the Gribov multipliers $v^{(0)}({\bf x})$ and $v^{(0)}({\bf x})^{-1}$ put out each other in ${\hat A}^{(0)}_k$. forming, therefore, the group ${\cal G}_0$ of physical values. 

\medskip Thus we have proven the equivalence of the notions ``topologically trivial'' and ``Poincare invariant'' for YM fields ${\hat A}^{(0)}_k$, forming, therefore, the group ${\cal G}_0$ of physical values. And then, in the YM model with vacuum BPS monopoles  quantized by Dirac \cite{LP1,David3,Pervush2,LP2,fund}, the reduction of the decomposition (\ref{direct}) to the reduction by the topologies $n\in{\bf Z}$ (i.e. by the YM modes ${\hat A}^{(n)}_k$) becomes also obvious.

\bigskip Very often in theoretical physics one deals with restricted operators (in particular, YM fields are such fields).  When such operators form a Banach $*$-algebra in a (quasi)Hilbert space $ \sigma$, Eq. (\ref{sa}) gives the decomposition
\be \label{pipi}
\pi=\oplus_{\tau\in {\cal F}}~~ \pi^{(\tau)}
\ee
of the representation $\pi$ of the observables algebra $\cal U$ to the subrepresentations $\pi^{(\tau)}$ in the subspaces transformed by a  representation multiple $\tau$  of the group $\bf \Gamma$. The set ${\cal F}$ entering Eq. (\ref{pipi}) is called {\it the physical sector} of the group $\bf \Gamma$. Herewith the vacuum representation $\pi_{\rm vac}$  of the algebra $\cal U$ is contained, obviously, in the representation $\pi_0$ corresponding to the trivial  representation $\tau_0(h)\equiv 0$ of the group $\bf \Gamma$. The states corresponding to the vectors from the spaces with $\tau\neq \tau_0$ were called {\it charged states} in the monograph \cite{Logunov}.

\medskip As it was mentioned already in Introduction, this construction of the ``non-vacuum sector'', involving the physical Hilbert space $\cal H$ and the  decompositions (\ref{direct}) and (\ref{pipi}), is suitable for the description of exitations, multipoles, over the YM-Higgs BPS monopole vacuum \cite{LP1,David3,Pervush2,LP2,fund} in the Dirac ``fundamental'' quantization scheme \cite{Dir}  applied.

\bigskip The following statement \cite{Logunov} is correct for the representation $\pi$, (\ref{pipi}).

{\it Let us suppose that the following natural technical supposition is fulfilled: for any elements $A$ and $B$ from the algebra $\cal B$ of physical values the expression $<\vert\gamma_h(A\alpha(a,1)(B))\vert>$ as a distribution in $a\in {\bf M}$} ({\it $M$ is the Minkowski space}) {\it depends continuously on $h\in \bf \Gamma$} ({\it as on a parameter}). {\it Then:}

(a) {\it the symmetry of the algebra $\cal B$ with respect to the Poincare group ${\cal B}_0$ is realized unitary in $\cal H$, herewith the spectrality condition \footnote{This means \cite{Logunov} that there exists the complete system of states in the Hilbert space $\cal H$ with the nonnegative energy.} is fulfilled};

(b) {\it If $\Psi_0$ is a unique} ({\it to within a multiplier}) {\it translation invariant vector in in $\cal H$, then the representation $\pi$ of the algebra $\cal B$ is unreducible}. 

\section{Specific the ``discrete''violation of the $U(1)$ gauge symmetry.}
The specific of gauge theories is in the question about the symmetry with respect to gauge transformations of the algebra $\cal B$ of physical (gauge invariant) values.

Let us consider now an Abelian gauge model involving the $U(1)$ gauge group and ``minimal''  coupling of the gauge field with matter fields. Suppose that the ``small'' gauge group ${\cal G}_0$ is the same as in the previous section. The role of the ''large'' group is played by ${\cal G}={\cal G}_0\times {\bf \Gamma}$, where the (compact) effective gauge group $\bf \Gamma$ is generated by the $U(1)$ subgroup and, perhaps, by some subgroup $H$. Herewith the $U(1)$ gauge subgroup is associated with the second kind gauge invariance in the considered model \footnote{In QED/Maxwell electrodynamics, for instance, the second kind gauge invariance means the invariance of the Maxwell equations with respect to the gradient transformations of the 4-potential $A_\mu$:
$$ {\bf A}\to {\bf A}+{\rm grad} f; \quad \varphi \to \varphi -\frac 1 c \frac{\partial \varphi}{\partial t}, $$
with $f(x,y,z,t)$ being an arbitrary function of the coordinates $(x,y,z)$ and the time $t$ \cite{Ger}.
} acting with the authomorphisms $\gamma_h$ ($h\in \Gamma$) of the shape
\be \label{isom}
\gamma_h(X)=e^{iqX_c} X \quad {\rm at}\quad h=e^{iec}; \quad c\in{\bf R}
\ee
 (with $q$ being the charge which the local quantum field $X\in{\cal F}$ carries) on these quantum fields $X$. 
 
 As we have discussed in the previous section, the authomorphisms $\gamma_h$ leave invariant a state $s$ of the algebra $\cal B$ and can be realized \cite{Logunov} by the  unitary operators $V(h)$ in the physical Hilbert space $\cal H$. 
 
 We shall speak that the gauge $U(1)$ symmetry {\it is broken down spontaneously} if the group of operators $V(h)$ ($h\in U(1)$) is not contained in the  von Neumann algebra \footnote{Let the algebra ${\cal B}({\cal H})$ of all the linear restricted operators be given over a Hilbert space $\cal H$. Then the topology on ${\cal B}({\cal H})$ that is specified with the seminorms 
$$ p_{\Psi_1,\dots, \Psi_n}^{\Phi_1,\dots, \Phi_n}(A)={\rm max}_{j=1,\dots,n} \vert <\Phi_j, A\Psi_j> \vert$$
 is called \cite{Logunov} {\it the weak topology} ($W$-{\it topology}). Here $n$ is a natural number while $\Phi_1,\dots, \Phi_n$ and $\Psi_1,\dots, \Psi_n$ arbitrary vectors in $\cal H$. 
 
 In this terminology, any involutive subalgebra (with the unit element) in ${\cal B}({\cal H})$ closed in the $W$-topology is called {\it the von Neumann algebra} \cite{Logunov}. 
 
\medskip Also the following important definition \cite{Logunov} will be necessary for our further consideration. 

In definition, the {\it commutant} of the subset $\textsl{R} \subset {\cal B}({\cal H})$ is the set  $\textsl{R}^{c}$ of those operators from $A\in {\cal B}({\cal H})$ that commute with all the elements from  $\textsl{R}$. It is obwious that a commutant is an algebra. If the set $\textsl{R}$ is closed with respect to the conjugation operation (i.e. $A^*\in \textsl{R}$ if  $A\in \textsl{R}$), than the commutant $\textsl{R}^{c}$ is an involutive subalgebra in ${\cal B}({\cal H})$.

Side by side with the notion ``commutant'' given above, one can give the notion {\it bicommutant}, that is \cite{Logunov} $\textsl{R}^{cc}\equiv (\textsl{R}^{c})^{c}$.

It turns out \cite{Logunov,Najm} that {\it the weak operator closure of an arbitrary involutive algebra $\cal U$ in ${\cal B}({\cal H})$ coincides with the bicommutant ${\cal U}^{cc}$}. {\it In particular, any von Neumann algebra coincides with its bicommutant}: ${\cal U}^{cc}={\cal U}$

In this is the essence of the so called {\it von Neumann density theorem.}
} $\pi({\cal B})^{cc}$ of physical values (and whence also in the von Neumann algebra $\pi({\cal U})^{cc}$ of observable values). The operator of the complete charge in such a theory, being no observable value, has a purely fictious sense and cannot be represented in a  reasonable sense as the integral $\int {\cal J}^0(x)d^3x$ from the zero current's component (since a current in an Abelian gauge theory is an observable field). 

In the case of an irreducible  representation of the algebra $\cal B$ of physical values, we shall consider the  symmetry of this algebra $\cal B$ with respect to the group of the $*$-automorphisms as a spontaneously no unbroken; herewith these  automorphisms are realized via an unitary representation of this group. In the case when the representation $\pi$ of the algebra $\cal B$ is reducible, one can impose an additional restriction onto the definition of a symmetry broken down spontaneously: the operators of the group representation should belong to the von Neumann algebra of physical values. 

The picture of the spontaneous break down of a gauge symmetry in Abelian models on the level of the unphysical quasi-Hilbert space ${\Huge G}'$ discovers a considerable resemblance with that occurring in the Goldstone theory (see, for instance, \S 10.3 in  \cite{Logunov} or \S 5.3 in \cite{Cheng}); in particular, the matrix elemnts of the current, $<\Phi\vert\tilde j^\mu (p)\vert 0 >$ possess singularities at $p^2=0$ which can be interpreted as the presence of a Goldstone boson in the considered Abelian gauge model. However, this resemblance is purely formal and this  Goldstone boson disappears at going over to the physical representation. 

In this is just the one of the signs of the Higgs mechanism, i.e. the effect acquiring  the mass by the gauge vector field at the spontaneous break of the gauge symmetry (by means of the ``absorption of the Goldstone boson'').

 Since we are interested, first of all, in the  spontaneous break down of the gauge $U(1)$ symmetry, then concerning the second (possible) symmetry group $H\subset {\bf \Gamma}$, we shall suppose that the symmetry with respect to $H$ is not broken down and that the automorphisms $\gamma_h$ ($h\in H$) of the field algebra $\cal F$ are realized with the quasi-unitary operators in ${\Huge G}_1$ which maintain invariant the vacuum vector $\vert 0>$.

\medskip At thse assumptions the following two suppositions \cite{Logunov} are correct.

\medskip {\bf Supposition 1.} {\it Let the} ({\it compact}) {\it effective group $\bf \Gamma$ be generated by the subgroups $U(1)$ and $H$} ({\it and the above assumption about $H$ is made}) {\it while the symmetry of the algebra $\cal B$} ({\it of physical values}) {\it with respect to the gauge group $U(1)$ is broken down spontaneously}. {\it Then}:

(a) {\it the group of automorphisms} $\gamma_h$ ($h\in U(1)$) {\it of the field algebra $\cal F$ } ({\it in a local Lorentz-covariant  gauge}) \footnote{For a $\xi$-gauge this means \cite{Logunov} $\partial _\mu A^\mu (x)=\xi \Lambda (x)$, with $\Lambda (x)$ being a quasi-Hermitian scalar field satisfying the d'Alembert equation
$$ \partial_\mu \partial^\mu  (x)=0$$
and the trivial canonical commutation relations
$$ [\Lambda(x),\Lambda(y)]=0. $$
 } {\it is not realized with the group of} {\it quasi}-{\it unitary operators in the} {\it quasi}-{\it Hilbert space ${\Huge G}_1$ maintaining invariant the vacuum vector $\vert 0>$};
 
 (b) {\it There exists an element $X$ of the polinomial field algebra $\cal P({\bf M})$ such that}
 
 \be \label{10128}
 <0\vert D(X)\vert 0>\equiv <0\vert i\int_{x_0={\rm const}}[j^0(x), X]d^3x \vert0>\neq 0.
 \ee
 
 To prove the first statement in Supposition 1, let us assume that the group of automorphisms $\gamma_h$ ($h\in U(1)$) of the field algebra $\cal F$  is  {\it realized} with the group of quasi-unitary operators ${\cal U}(h)$ in ${\Huge G}_1$ maintaining invariant the vacuum vector. Then it is easy to see that $<0\vert\gamma_h(X)\vert 0>=<0\vert X\vert 0>$ at $X\in {\cal F}$, $h\in U(1)$. Besides that, it follows from the conditions of Supposition 1 that the same equality is correct for $h\in H$; this means  it is correct at all $h\in {\bf \Gamma}$. As a result, the state (\ref{sa}) on $\cal B$ is $<0\vert s(x)\vert 0>$. One can show \cite{Logunov} that the vacuum is not deheratated in the physical Hilbert space $\cal H$, that the representation $\pi$ of the algebra $\cal B$ in $\cal H$ is irriducible and that the gauge group $U(1)$ is realized with  unitary operators $V(h)$ in $\cal H$. These  unitary operators can be set with the formula \cite{Logunov}
 \be \label{1032}
 V(h)[\Phi]=[{\cal V}(h)\Phi]\quad {\rm at}\quad \Phi\in {\Huge G}_1.
 \ee
 Here the map $h\to {\cal V}(h)$ sets the (quasi)-unitary representation of the group $\bf \Gamma$ in ${\Huge G}_1$.
 
 Since the representation $\pi$ is irriducible, this means that the gauge symmetry group $U(1)$ is not broken down spontaneously. The just obtained contradiction with the conditions of  Supposition 1 proves (a).
 
 The second statement can be proven in a similar way. Really, if one suppose that $<0\vert D(X)\vert 0>=0$ for all $X\in {\cal P}({\bf M})$, then also $<0\vert\gamma_h(X)\vert 0>=<0\vert X\vert 0>$ for all $X\in {\cal P}({\bf M})$, $h\in U(1)$; since the ``large'' (initial) gauge symmetry group $\cal G$ acts continuously on $\cal F$, the obtained equality is correct for all $X\in {\cal F}$, $h\in U(1)$. Treating further with the matter as at proving the previous statement, we find that the gauge  group $U(1)$ again is not broken down spontaneously; this contradiction proves (b).
 
 \bigskip The on principle another situation arises in the case of the spontaneous break down of the $U(1)$ gauge symmetry in the ``discrete'' way (\ref{discret}). The thing is that (\ref{discret}) is an isomorphism; this means that now formally ``the group of operators $V(h)$ ($h\in U(1)$) is {\it contained} in the  von Neumann algebra $\pi({\cal B})^{cc}$ of physical values'', in contrast to the assumptions of Supposition 1. Figuratively speaking, at the isomorphism (\ref{discret}) the $U(1)$ gauge symmetry is broken down spontaneously from the ``physical'' point of view (herewith the continuous $U(1)\simeq S^1$ group space disintegrates into its topological sectors $n\in {\bf Z}$ slotted by domain wells \cite{disc}), while ``mathematically'' this $U(1)$ gauge symmetry remains unbroken.
 
 \medskip The said has a crucial importance in the various gauge models involving the violating the $U(1)$  symmetry. 
 
 Besides the liquid ${\rm He}^4$ model \cite{disc,Halatnikov,Kapitza,Landau}, as discussed in Section 1, the scheme (\ref{discret}) of violating the $U(1)$  symmetry can be applied in QCD in the string confinement model \cite{Nambu} (see also p. 371 in the monograph \cite{Cheng}). The key point of that model is, as it is well known, is the QCD vacuum consisting of (light) quark-antiquark pairs and gluons. Herewith quarks and antiquarks are connected with (infinitely) thin gluonic tubes (strings). In this is the specific of the {\it confinement phase} in the {\it dual Abelian theory} \cite{Hooft}.  In this phase quarks in the fundamental representation possess, as purely electric objects, electric charges $\frac e N$ and are connected by one vortex.
 On the other hand, Higgs modes, as objects dual to quarks and gluons, becomes purely magnetic objects. 
 
 This points onto two things necessary for the ``string'' confinement picture. Firstly, the initial $SU(3)_{\rm col}$ gauge symmetry group should be broken down to its $U(1)$ subgroup, say in the
 \be \label{MAG} SU(3)_{\rm col} \to SU(2)_{\rm col}\to U(1)\otimes U(1)\ee
 way via the standard Higgs mechanism \footnote{Eq. (\ref{MAG}) reflects fixing the so called {\it maximal Abelian gauge} (MAG), involving singling out the diagonal Gell-Mann matrices $\lambda_3$, $\lambda_8$ from the total set of the Gell-Mann matrices $\bf \lambda$.}. 
 
 Secondly, the appearance of (infinitely thin) "magnetic"  flux tubes (strings) confining quarks in QCD can be explained (and the author thinks that it is the only reasonable explanation of the effect) in the framework of the well-known {\it Nielsen-Olesen} (NO) model \cite{No} involving topologically nontrivial solutions, Nielsen-Olesen vortices.
 
 Going over to the ``theoretical group language'', the cause  appearing these NO vortices is concealed \cite{disc} in destroying the $U(1)$ [$U(1)\otimes U(1$] gauge symmetry, as in  (\ref{MAG}), in the 
 
 \be \label{disw}
 U(1)\to{\bf Z}
 \ee
 way.
 
Herewith we should keep in our mind that such breakdown of the "contineous" $U(1)$ gauge symmetry with NO vortices  \cite{No} appearing is a {\it first order} phase transition from the thermodynamics point of view (similar to that  occurring in Type II superconductors \cite{Abrikosov}). This implies (as it was supposed in \cite{disc}) gradual destroying  the "contineous" $U(1)\simeq S^1$ group space by "chipping-off" some topological sectors (with fixed integer numbers $n$) from this group space. In simultaneous coexisting the contineous and discrete spaces (latter is formed by means of {\it gradual} arising domain walls inside the former one \cite{disc}) is just  the essence of the first order phase transition taling place in the NO model  \cite{No}. 
 
\end{document}